\documentclass[runningheads]{llncs}
\usepackage{epsfig}
\usepackage{epstopdf}
\usepackage{newunicodechar}
\usepackage[utf8]{inputenx}
\DeclareUnicodeCharacter{2212}{-}
\DeclareUnicodeCharacter{2217}{-}
\DeclareUnicodeCharacter{00A0}{~}
\usepackage{cite}

\usepackage{amsmath}
\usepackage{pdfpages}
\usepackage{graphicx}
\begin{document}

\title{Deep Learning Algorithm for Threat Detection in Hackers Forum (Deep Web)}
%
%
\author{Victor Adewopo\inst{1}* \and Bilal Gonen\inst{1} \and Nelly Elsayed\inst{1} \and Murat Ozer\inst{1} \and Zaghloul Saad Elsayed\inst{1}}
\authorrunning{Victor Adewopo et al.}
%
\institute{University of Cincinnati, School of Information Technology, Cincinnati, Ohio, USA\\
\email{adewopva@mail.uc.edu}
}
\maketitle              

\begin{abstract}
In our current society, the inter-connectivity of devices provides easy access for netizens to utilize cyberspace technology for illegal activities.
The deep web platform is a consummative ecosystem shielded by boundaries of trust, information sharing,  trade-off,  and review systems. Domain knowledge is shared among experts in hackers forums which contain indicators of compromise that can be explored for cyberthreat intelligence. Developing tools that can be deployed for threat detection is integral in securing digital communication in cyberspace. In this paper, we addressed the use of TOR relay nodes for anonymizing communications in deep web forums. We propose a novel approach for detecting cyberthreats using a deep learning algorithm Long Short-Term Memory (LSTM). The developed model outperformed the experimental results of other researchers in this problem domain with an accuracy of 94\% and precision of 90\%. Our model can be easily deployed by organizations in securing digital communications and detection of vulnerability exposure before cyberattack. 
\end{abstract}

\section{Introduction}
In our contemporary society, the existence of interconnected devices in cyberspace is faced with high vulnerabilities. However, the ubiquitous use of cyber-technology, the internet of things, and smart technologies steered the revolution and advancement in the quality of human lives. Cyberspace is one of the most complex systems ever built by human beings, the resources are widely used by many, but only a few understand the complexities of cyberspace. The availability of sophisticated technologies enables criminals to exploit new ways of committing crimes. Criminogenic activities in cyberspace are a critical societal problem that requires the swift intervention of cybersecurity experts \cite{adewopo2019plunge,9457567}. There are often misconceptions with the terms "Surface web," "Deep web," and "Dark web," they are relatively interconnected but do not mean exactly the same thing. The surface web is web pages unencrypted and can be accessed using traditional search engines (e.g., Google, Bing, Yahoo). The surface web consists of billions of static web pages, and it occupies only 10\% of the internet space. 

The dark web is a layer within the deep web and is not accessible using a standard browser. They are intentionally hidden parts of the web but can be accessed using a specific URL address. The deep web contains about 90\% of the contents available on the internet \cite{Zhao2016}. Deep web contents are available on the web, but cannot be indexed or accessed by using regular search engines. Several techniques and tools have been designed to understand and crawl the deep web. A recent report indicated the low harvest rate of the deep web- about 647,000 distinct web forms was found by sampling 25 million pages using Google index \cite{Zhao2016, dragut2012deep}. Darknet is an encrypted network technology that uses the internet infrastructure and can only be accessed using special network configuration and software tools to access its contents \cite{Mansfield-Devine2009}.
\begin{figure}
    \centering
    \includegraphics[scale=0.5]{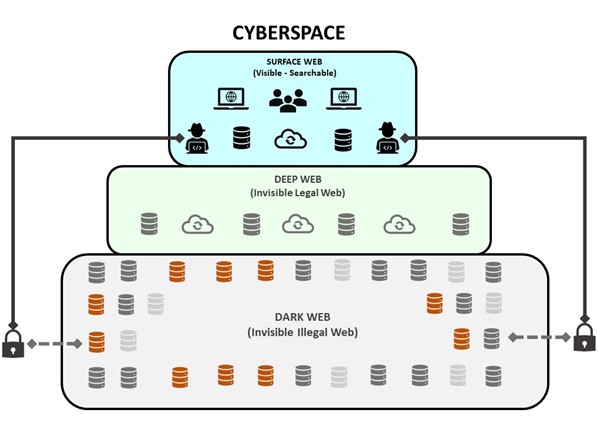}
    \caption{Layers of Internet: Surface Web, Deep Web, And Dark Web Explained}
    \label{fig:f1}
\end{figure}
Traditionally, deep web hacker forums are used for criminal activities, including different types of cybercrime, procurement of illegal drugs, child pornography, cryptocurrency markets, and hacking \cite{Robertson2017}. Cryptocurrency (i.e., Bitcoins) is adopted as a method of payment in hacker forums because of its' anonymous characteristics, which make it more difficult to link the users with any specific bank account details \cite{ozer2019prevention}. In the quest for high-quality security-related information, developing methodologies to crawl the deep web and excavate wealth of information for cyberthreat intelligence is a paramount concern \cite{adewopo2020exploring}. A Cybersecurity firm "UpGuard" reported 419 million Facebook users' data breach in April 2019. Data exposed publicly on the Amazon server includes; passwords, user IDs, and check-ins \cite{Perez}.
Similarly, in September 2018, over 50 million Facebook users' breached data was auctioned at a bitcoin value of 3\$ per each user's data on the Darknet \cite{Cuthbertson}. Nefarious actors have been able to build a digital ecosystem that provides platforms famously acclaimed to promote activities such as cybercrime, terrorism, hacking, procurements of illegal drugs, arms deals, and anonymous markets \cite{Guitton2013}. The hidden nature of Darknet websites aided fraudsters in utilizing cyberspace for criminal activities.

\section{User's Motivation In Deep web forums}
Why do people want to use the deep web? Over the years, the deep web has been traditionally used by actors to carry out nefarious activities. In countries with high censorship of all digital communications, including indiscriminate censorship of media by oppressive regimes have led to an increase in the use of hidden services. For example, in March 2019, a darknet website was used to publish the leaked emails and official documents of the Russian Government to expose corruption and promote transparency \cite{Mod}. 
The darknet is classified as a platform for private networks and infrastructure for the public systems, in that the infrastructure is considered as a public entity. When the services are used as a private platform, users communicate anonymously on the network, and they consent to store data on their local computers. Most of these stored data are inaccessible to the local users who harbor the stored data in their devices. Platforms acquire the existing infrastructure of public networks by enhancing security features and constraints to privatize communications \cite{Plantin2018}. Gehl et al. \cite{Gehl2019} suggest that the platform tends to be of a more private network while infrastructures lean more toward public entities. The anonymous nature of the darknet could provide individuals with both well-intended motives and malicious intentions to carry out various activities on the platform \cite{Gehl2019}. Wang et al \cite{Wang2018} analyzed the personal characteristics of the darknet users to identify typical individual user's profile. They aimed at developing a tool to track darknet users and reduce cybercrime. A three-step approach was used in analyzing the characteristics of darknet website users through Named Entity Recognition and Information Extraction.
A.	Block Filtration.
B.	Attribute candidate generation.
C.	Attribute candidate verification.
\subsection{Use of TOR for Anonymizing Services in 
 Deep Web Forums}
 TOR is an open-source software that secures anonymous communications and also allows users to host hidden services on the TOR network \cite{Tsai2006}. TOR services can dynamically allow users to access the contents of websites without revealing their identity, which prevents government censors from monitoring sites accessed by users. Accordingly, bloggers could host websites and hidden services by publishing contents on Distributed Hash Tables (DHT) and TOR services which makes the directory of the database random in which all participating nodes can access and retrieve information \cite{schafer2019blackwidow}. This makes it difficult to censor, control, or access the location of the website's contents \cite{Owen2015, Irudayaraj2011}. The downside of such services is that the ubiquitous utilization of hidden services could pose a significant threat compared to the benefits mentioned above. Hackers also use TOR services, and the lack of national laws and international cooperation among relevant agencies are two significant challenges in combating the investigation of criminal activities committed using Tor hidden services \cite{Everett2015}. TOR software ensures user's anonymity by relying on an overlay of networks using three relay nodes: (1) Entry relay nodes, (2) Middle relay nodes, and (3) Exit relay nodes.
\begin{itemize}
    \item Entry Relay Nodes
\newline The entry relay nodes serve as the entry point to TOR services, also called entry guards, which receive the TOR traffic and pass it onto other nodes that anonymize users as the originator of traffic. This node encrypts communication between the traffic source and the entry relay nodes. The IP addresses of the users connected to the TOR network are logged in the list of guard nodes to keep track and distinguish entry relay nodes \cite{Chaabane2010, Elahi2012}
\item Middle Relay Nodes 
\newline The middle relay nodes transport traffic from an entry relay to the exit relay to ensure anonymity and bridge gaps between the entry and exit relay. Each intermediate node can only communicate with its predecessor and successor relay in Tor services. The nodes are configured in such a way that communications cannot be linked (send/receive data) without the intermediate nodes \cite{adewopo2019plunge}. 
\item Exit Relay Nodes
\newline The exit relay is the final relay that Tor traffic passes through before it reaches its destination. The IP address of the exit relay is interpreted as the source of TOR traffic. The exit nodes contain the key of encrypted information in the Tor network. The original information from the source traffic can only be recovered through the exit relay.
\end{itemize}
The exit relay servers are often the sources for complaints and associated with illegal activities perpetrated by source traffic, which attracts law enforcement agencies' attention. Sun et al. \cite{Sun} discovered two ways of observing the TOR network and identifying anonymous users, which could be achieved by either owning multiple TOR relays or manipulating the underlying network communications. TOR is the most common anonymizing service that runs over seven thousand relays. TOR uses bridges, which are new TOR routers that are not listed on the Tor network. Some Internet Service Providers (ISPs) blocked TOR nodes and access to the TOR network by filtering the IP address of TOR servers. Bridges are used in circumventing blocked IP addresses of listed Tor relays \cite{Chaabane2010, Tsai2006}

\section{Pattern of Cyberthreat between 2019 and 2021}\label{sec:Pattern of attack}
A cyberattack
can come in the form of malware that is virulent in nature. Oftentimes, they may look benign on sight but are usually embedded with malicious infections. Most malware infections originate in a form of attachment with the ability to hide, replicate themselves on a target system, and spread to other computers on a networked device, while some malicious infections such as the trojan virus do not replicate themselves. Malware is somewhat platform-centric, targeting specific security loopholes to explore areas with high vulnerability. In a malicious ransomware attack, victims' file disks are usually partitioned, changed, or encrypted with a key that can only be decrypted by a mathematically related key ~\cite{Upadhyaya2017}.
The existence of sophisticated technology can be deployed in predicting cyberattacks before the actual occurrence. The Wall Street Journal reported the likelihood of a federal judiciary’s systems breached in a SolarWinds hack which prompted the federal court to store sensitive documents filled in a stand-alone computer system \cite{volz_mcmillan_2021}. In March 2021, Over 30,000 organizations across the U.S were hacked by cyber espionage focused on siphoning email communications through the Microsoft exchange flaws. In 2019, IBM security reported that the average cost of a data breach in the U.S. is about \$3.9 million, the average life cycle of a data breach stands at 314 days, and the impact on an organization can span over 4 years with 67\% of the cost occurring within the first year \cite{PonemonInstitute2019}. In 2020, over 273\% increase in the number of breached records in the first quarter of 2020 was observed as compared to the first quarter of 2019 \cite{dutta}. 
In July 2020, attackers swindle \$121,000 worth of bitcoin in 3,000 transactions, by hacking Twitter accounts of 130 high profile US personalities and resetting 45 user accounts passwords. The phone Spear-phishing method of attack was targeted towards a few employees to gain access to internal systems. 
In April 2020, approximately 500,000 stolen zoom passwords were available for sale in darknet markets while some account credentials were made available free. Information compromised includes login credentials, victim personal meeting URLs, and host keys~\cite{dutta}.
Nefarious actors leverage the virtual ecosystem of cyberspace to perpetrate sophisticated cyberattacks launched against an unsuspecting individual. 
In April 2019, there were two incident reports of Facebook data breach. A database with about 419 million unique Facebook users' record, containing ID, phone numbers, and other personal details linked to users account was exposed on an Amazon cloud service. Another data breach contains Users' ID, Facebook friends, likes, photos, Facebook check-ins, and some other sensitive information that was leaked in the Amazon S3 bucket cloud server~\cite{cisomag}.
In June 2019, US medical bill debt collector, American Medical Collection Agency (AMCA) filed for bankruptcy after eight months-long data breach, and huge expenses (\$2.5 Million loans) suffered when their system was hacked that led to the exposure of about 20 million Americans' personal information \cite{okazaki2020understanding}.
In July 2019, Capital One security breach leaked over 140,000 social security numbers of their clients, personal information, and around 80,000 bank account numbers were exposed \cite{okazaki2020understanding}. The breach is estimated to cost the firm about \$100 - \$150 million.
In November 2019, personal data belonging to 1.2 billion unique people were exposed on an Elasticsearch server that contains more than four terabytes of data \cite{cisomag,okazaki2020understanding}.
\subsection{Number of Records Compromised and Types of Data Breach (2005-2019)}
\begin{figure}[!htbp]
  \centering
  \includegraphics[width=1.0\linewidth]{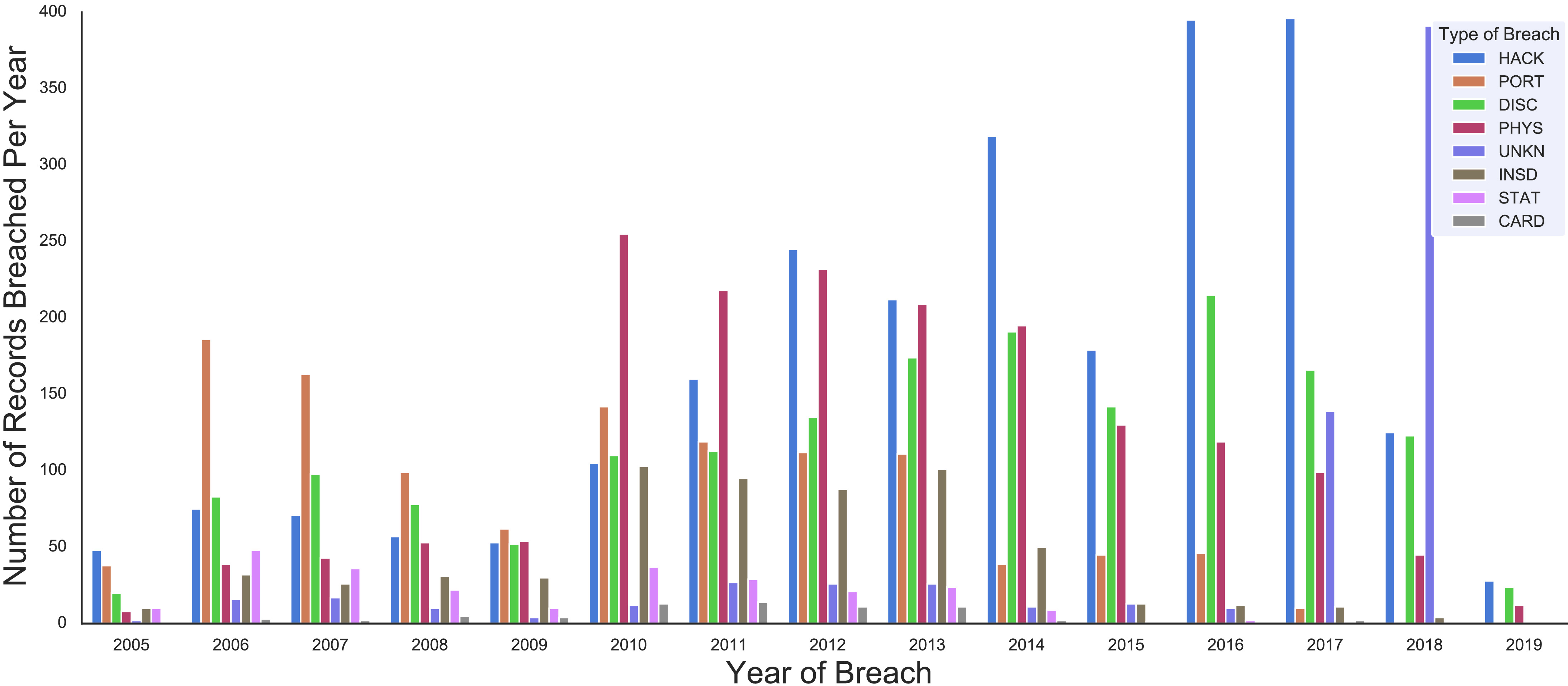}
   \caption{Total Number of Compromised Records in Millions (2005-2019)}
 \label{fig:BreachperYear-06}
  \end{figure}

In this research, we analyzed the Privacy Rights Clearinghouse (PRC) dataset containing the report of data breaches for over 14 years.
The dynamic nature of the technology industry creates security loopholes that can be explored for cyberattacks, some organization suffers a major loss, and often fold up due to unprecedented events. Most security and data breaches span across all major industries. The Privacy Rights Clearinghouse (PRC) Chronology of Data Breaches reported more than ten billion records breached. Figure (\ref{fig:BreachperYear-06}) discloses that unauthorized access by a non-member of an organization (HACK) led to the highest records breach with over one billion data compromised in both 2016 and 2017. It is also noteworthy that the pattern of attacks changed from physical access to information and portable devices to unauthorized access to information (i.e. Malware, DDOS) and unintended disclosure of information between 2005 to 2012 and 2013 to 2019. Most cyberattacks are launched through the use of sophisticated technology; this is also supported by the research work conducted by Decary et al. \cite{decary2017police}.

\subsection{Occurrences of Cyberattacks Per States in the US}
\begin{figure}[!htbp]
\centering
\includegraphics[width=0.8\linewidth]{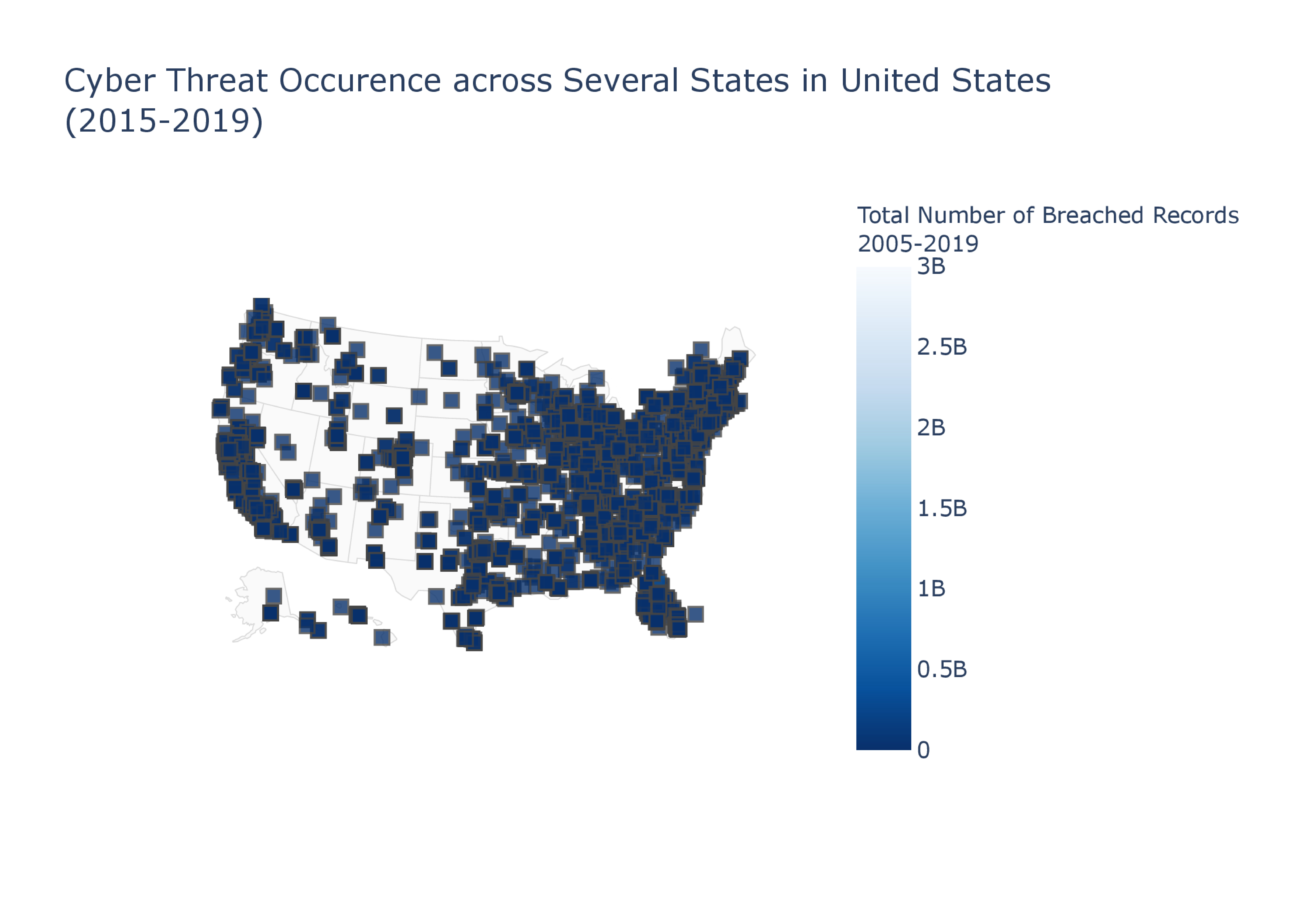}
\caption{US Map Showing Numbers of Breached Records Across Various Cities in US (2005-2019)}
\label{fig:Finalnewplot-09}
\end{figure}

Several states in the US have started paying more attention to secure vital information and also invested in improving data protection. The New York Penal Law S. 156.10 punishes the act of knowing use of a computer to gain unauthorized access to material, otherwise known as computer trespass. It was reported that 31 states in the US enacted cybersecurity legislation that significantly tackles issues relating to cybersecurity threats\cite{fonseca2017cybersecurity}, some states also have specific laws and policies regarding data breach. Most states have enacted legislation that binds private or governmental entities to notify individuals of security breaches of information involving personally identifiable information after a certain period of days.
Figure~(\ref{fig:Finalnewplot-09}) further illustrates the total number of records breached in different parts of the United States. The total number of records breached indicated across the bar-line is measured in billions with clusters in South and Northeast of US which indicates a higher risk of data breach for organizations located in the southeast and northeast. There is only a few research that has taken into consideration the crime distribution rate in the united states using geo-spatial tools in identifying the pattern of crime ad type of data breach. The total number of records breached in different states ranges from zero to five hundred million data. Only a few states have the total number of records breached between one billion and two billion reports of data breach. The research of Khey et al. \cite{khey2013examining} focused on the spatial distribution of data breaches in the United States and risk profiling of vulnerabilities across geographical locations.
\section{Using the Deep Web (Hackers Forums) as a  Source of Cyberthreat Intelligences}
In the era of big data, crawling the deep web is very important to get insights into information embedded in the deep web; a large amount of data exists in cyberspace and can be accessed through various interfaces. The inability of search engines to crawl the deep web limits the information provided to users. Oftentimes, users have to spend extra time and effort querying a website manually to access the contents of a website. In their study, Dixit et al.~\cite{Manvi2013} developed an Ontology-based adaptive crawler for the hidden web by mapping relationships between webpages using four components (Ontology Builder, Hidden Web Miner, Result Processor, and Domain-Specific - Flow Diagram). The research of Dong et al.~\cite{Dong} utilized a lightweight framework to predict cyberthreats from darknet data. The method applied a scrappy crawler with proxy VPN to parse the web page for relevant information which is processed through verification techniques to generate cyberthreat warnings. Azene et al. \cite{Zenebe2019} utilized descriptive and predictive analytic machine learning techniques on a data-set from darknet forum to discover valuable cyberthreat intelligence. IBM Watson analytic tools and WEKA was used to classify types of exploits targeted by hackers in the forums. Their results showed that password cracker, key logger, and remote administration tools are the most frequent tools targeted by influential authors in the forum 
Research work in \cite{Kadoguchi2019, Robertson2017} adopted a framework in predicting the future occurrence of a cyberattack by mapping different data-sets from the dark web, socio-personal and technical indicators with online cybersecurity report forums. A combination of human and automated techniques was used to extract live data from the dark web, insights gathered from online forums and community can be used to predict malicious acts before they occur. The use of search engines and spider services on the TOR network was used to collate a list of more than 100 malicious hacker websites. 

\subsection{Extracting information from the Deep web (Hackers Forum)}
The deep web is unindexed, making it challenging to analyze the contents or have access to information on darknet websites. Standard search engines cannot harvest or crawl the contents of darknet websites. Two methods can be used to crawl the contents of darknet websites. The first is through prior knowledge base and queries, and second, a non-prior knowledge base with the use of greedy queries. The attempts to close darknet websites always result in a ripple effect. 
One of the major challenges encountered in crawling the deep web forums was CAPTCHA authentication, username and password, invitation code, and filling of query forms for automatic crawlers. The prototype system built-in \cite{He2013} tackled the challenge of query generation, empty page filtering, and URL duplication.
In this research, we utilized an open-source deep web dataset that contains discussion forums from two darknet markets (Silkroad \& Wall street) extracted from the Arizona State University database \cite{du2018identifying}. The data set contains over 128,000 posts from different discussion threads. Discussions were organized in a thread topic, and other users initiate discussions based on the thread title. 
The thread title is related to :
\begin{enumerate}
    \item Carding: Carding thread forums involve discussions on the trade of stolen credit card details, identity theft, and currency counterfeiting. The forums allow both vendors and customers to trade credit card details, bank account, and other personal information online.
    \item Newbie: This is a discussion thread for new users and inexperienced hackers. The discussion thread focuses mainly on welcoming new members, introducing basic concepts and rules of engagements. Some other forums limit new users to pending when they make payments to access advanced features.
    \item Scamming: Discussion in this threat is centered towards stealing of information, fraudulent activities, and transactions. One major characteristic of this thread is dishonest transactions and the listing of fake products.
    \item Hacking: Hackers forums consist of a wide variety of discussions ranging from software leak, malware, credit cards, data breach, identity theft, and financial information.
    \item Review Thread: The review thread in the forums represents a town hall meeting where both vendors and customers discuss a variety of topics and share knowledge base with each other. This thread also serves as a tool/page to report users who have violated the forum's policy or scammed other members of the forum.
\end{enumerate}  

\subsection{Deep Learning Algorithm for Threat Detection}
Deep learning is a subset of machine learning methods based on neural networks with representation learning. Deep Neural Network (DNN) is an artificial neural network with multiple layers between the input and output layers.
Machine learning algorithms can be supervised, semi-supervised or unsupervised.
The Long Short-Term Memory (LSTM) neural network is a special variant of Recurrent Neural Networks (RNN), which overcomes stability bottlenecks encountered in traditional Recurrent Neural Networks \cite{mohan2019compressed}. RNN is a class of Artificial Neural Network (ANN) that allows the use of previous outputs as the current input with hidden layers. This special type of ANN makes use of sequential and time-series information, RNN allows output from previous steps to be used as input to the current step while having hidden states. The unique feature of RNN is the sharing of parameters across each layer of the network, RNNs share the same weight parameter within each layer of the network and are adjusted through the processes of backpropagation and gradient descent. RNN stores information in memory because their current output is dependent on the previous computations. However, the major problem associated with RNNs are vanishing and exploding gradients, LSTMs were designed to overcome the bottlenecks of RNN by introducing new gates which allow better control over the gradient flow and enable better preservation of long-range dependencies \cite{bouktif2018optimal,Sylvia}.
The LSTM architecture is different from other deep learning architectures, LSTM model contains the memory cell and gates which are essential to the architecture namely; the input gate, output gate, and the forget gate. The LSTM regulates the flow of training information through these gates by selectively adding information.

\begin{align}
    i_t &= \sigma (W_i \cdot[h_{t-1}, x_t] + b_i)  \\
    O_t &= \sigma (W_o \cdot [h_{t−1}, x_t] + b_o)\\
    f_t &= \sigma (W_f \cdot [h_{t-1}, x_t] + b_f )\\
    C_t &= f_t ∗ C_{t−1} + i_t ∗ \tilde{C}_t\\
    h_t &= O_t ∗ tanh (C_t)\\
    \tilde{C}_t &= tanh (W_C\cdot [h_{t−1}, x_t] + b_C )   
\end{align}

\begin{figure}[!htbp]
    \centering
    \includegraphics[width=1.0\linewidth]{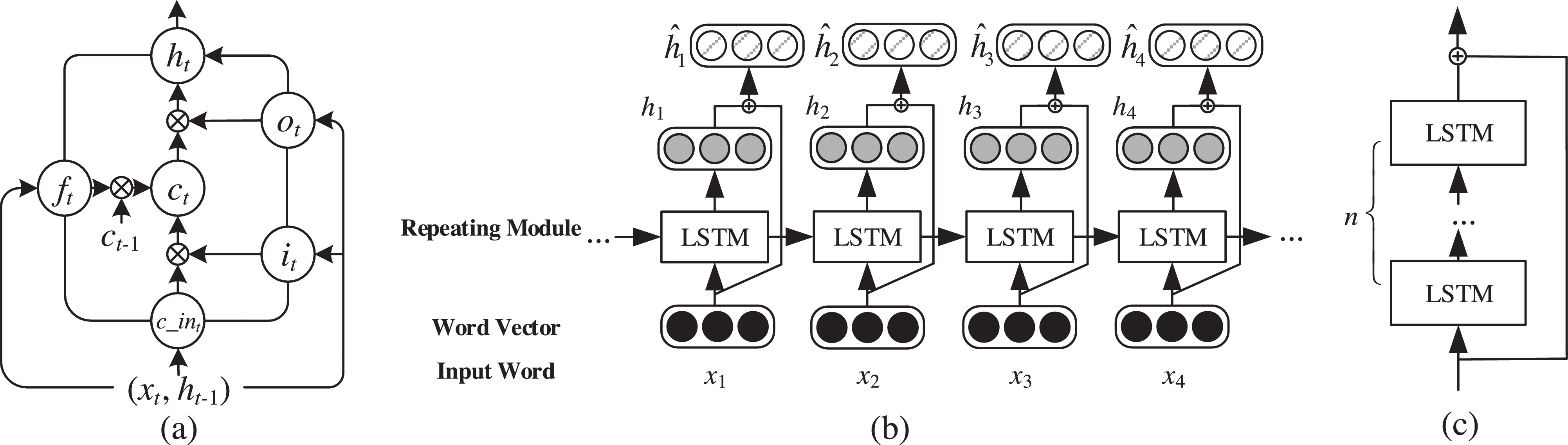}
    \caption{Simple illustration of LSTM cells}
    \label{fig:f6}
\end{figure}

The gates in LSTM cell enable it to preserve constant error that can be backpropagated, the input gate is represented by \(i\), output gate by \(o\), and forget gate by \(f\). The memory cell represented as \(C\) accumulates the state of information, and the cell output is given by \(h\), while the cell input at time \(t\) is denoted as \(x_t\). \(W\) are the weights for each of the gates and \(\tilde{C}\) is the updated cell state. These states are propagated ahead through the network, the forget
gate plays a crucial role in reducing over-fitting by not retaining all information from the previous time steps. See figure(\ref{fig:f6})\\

The performance of machine learning methods is heavily dependent on the choice of data representation (features) to which they are applied. In this study, we utilized LSTM to train the deep web data because LSTM performs better when training a large time-series corpus.  In this research, we curated a unique features extraction pipeline to extract time-series data from hackers' forums. The extracted features served as the input of the first hidden layer of our LSTM model architecture.
\section{Results and Discussion}
We used NER for features extraction and tokenizers to convert raw text to matrices. The average length of each text in the dataset is 150. The maximum word length is a sequence of 250 words. The input texts were padded to maintain a uniform sequence length.
The model was trained on twelve epochs with an accuracy of 91\% on train and 87\% on validation
data.
\begin{figure}[!htbp]
    \centering
    \includegraphics[width=0.6\linewidth]{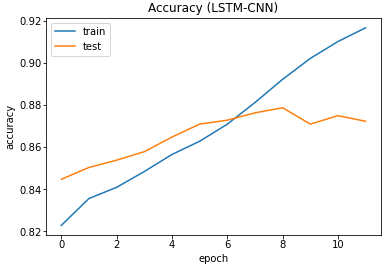}
    \caption{Accuracy of LSTM Model}
    \label{fig:f8}
\end{figure}\\
Figure (\ref{fig:f8}) shows the accuracy of the trained LSTM model on both train and validation data. 
We compared the performance of the developed model with other related work in predicting cyberthreat and identifying vulnerability exposures using deep web data. See table (\ref{table 5}).  The research of Dong et al. utilized a lightweight framework to predict cyberthreats from darknet data. The researchers monitored eight darknet marketplaces for a period of five months. Scrappy crawler was used with Polipo and Vidalia proxies to scrape information related to cybersecurity. Their model was able to predict 145 existing threats, 35 new threats, and newly developed hacking tools in the darknet marketplace.
The researchers experimented with SVM. SVM has the highest accuracy of 81\% and precision of 90\% in detecting threats \cite{Dong}.
The research of Azene et al. \cite{Zenebe2019} on cyberthreat discovery from the dark web utilized descriptive and predictive analytic machine learning techniques to discover valuable cyberthreat intelligence. Their classification model with Naive Bayes had an accuracy of 65\%  on web exploit, Random Tree 84\% on system exploits, and Random forest classifier has the highest accuracy of 97\% on system exploits.
The research of Ashok et al.\cite{Deb2018} developed a novel approach for predicting cyberthreat leveraging on hackers' sentiments. The researchers analyzed over four hundred thousand posts in different hacker forums. Their experimental results in organizations malicious email had the precision of 76\% and recall of 63\%.
The deep learning algorithm (LSTM) developed in this research outperforms every other base model algorithm reported in related research. As showcased in \ref{table 5}), LSTM had an accuracy of 94\%, precision of 90\%, and a recall of 91\%, which is the best result of deep learning algorithms in identifying anomalous cyberthreat texts and predicting vulnerability exposures leveraging on the discussion in hackers forum.
\begin{table*}[!htbp]
  \caption{Comparison of other baseline models}
  \label{table 5}
  \small\sf\centering
  \scalebox{0.8}{%
  \begin{tabular}{|p{35mm}|p{35mm}|p{20mm}|p{20mm}|p{20mm}|p{20mm}|}
    \hline
     Models&Algorithm & Accuracy& Precision & Recall&F1/ROC\\
\hline
My Model & LSTM
& 0.94&0.90&0.91&0.91 \\
\hline
&Random Forest (RFC)
& 0.80&0.95&0.95&0.75 \\

\hline
Dong et al.~\cite{Dong} & SVM
& 0.81&0.90&-&- \\
\hline
Azene et al\cite{Zenebe2019} &Naive Bayes
&0.65&-&-&0.85 \\
\hline
&Random Tree
& 0.84&-&-&0.71 \\
\hline
&Random Forest
& 0.97&-&-&0.91 \\
\hline
Arora et al. \cite{arora2019detection} &Random Forest(RFC)
& 0.80&0.81&0.80&0.79 \\
\hline
  \end{tabular}}
\end{table*}
\section{Conclusion}
Cyberspace is a global environment where innovation, creativity, and access to information have almost no limit. Most businesses rely on web-based applications such as Zoom, WebEx, Microsoft teams, and other communication platforms. Actors with malicious intent could exploit the vulnerabilities of these systems to steal personal information and engage in illegal activities. The occurrence of Cyberattacks has a significant impact on organizational activities and people's privacy. Based on the increased rate of occurrence gathering social and technical indicators useful for threat intelligence from hackers forum is imperative in developing tools for detecting cyberthreats before the actual occurrence. The model developed in this research can be deployed by security analysts and law enforcement agencies in identifying existing cyberthreats and predicting potential vulnerability exposures using deep learning algorithms. The results of our experiments indicate that our methodology is more effective in detecting cyberthreats with 90\% of accuracy.
This paper also identifies potential areas for further research in developing a methodology to scrape information from deep web forums securely using automatic smart crawlers.
As part of our future studies, we will use geospatial analysis in profiling risk factors of cyberthreat in various regions in the United States to establish correlations between deep web and surface web virtual ecosystem in orchestrating criminal activities by nefarious netizens.

\section*{Acknowledgment}
This material is based upon work supported by the National Science Foundation under Grant No. (CNS-1801593). Any opinions, findings, and conclusions, or recommendations expressed in this material are those of the author(s) and do not necessarily reflect the views of the National Science Foundation.

\bibliographystyle{splncs04}

\bibliography{mybibfile}

\end{document}